\documentclass{INTERSPEECH2024}
\usepackage{algorithm}
\usepackage{algpseudocode}
\usepackage{subcaption}
\usepackage{pgfplots,multirow}
\usepackage{url,hyperref}
\usepackage{stmaryrd}
\usepackage{pifont}
\pgfplotsset{compat=1.17} % Use this to ensure you're using a version of pgfplots that matches your LaTeX distribution
\usepackage{siunitx}
\interspeechcameraready 

\title{Exploring the Benefits of Tokenization of Discrete Acoustic Units}
\name{Avihu Dekel, Raul Fernandez}
\address{IBM Research}

\email{avihu.dekel@ibm.com, fernanra@us.ibm.com}

\newcommand{\cmark}{\ding{51}}%
\newcommand{\xmark}{\ding{55}}%

\begin{document}
\maketitle
\begin{abstract}
Tokenization algorithms that merge the units of a base vocabulary into larger,
variable-rate units have become standard in natural language processing tasks.
This idea, however, has been mostly overlooked 
when the vocabulary consists of phonemes or Discrete Acoustic Units (DAUs), an audio-based representation that is playing an increasingly important role due to the success of discrete language-modeling techniques.
In this paper, we showcase the advantages of tokenization of 
phonetic units and of DAUs on three prediction tasks: 
grapheme-to-phoneme, grapheme-to-DAUs, and unsupervised speech generation using DAU language modeling.
We demonstrate that tokenization yields significant improvements 
in terms of performance, as well as training and inference speed, across all three tasks. 
We also offer theoretical insights to provide some explanation for the superior performance observed.
% train a highly capable speech language model that exploits these insights, 
%Tokenization is widely used in natural language processing.
%However, it is not commonly applied in discrete acoustic units (DAUs).
%In this paper, we explore the advantages of tokenization on DAUs.
%Specifically, training models on tokenized DAUs yield significant performance improvements.
%This comes along with great improvements in training and inference speed.
%We showcase the advantages G2P, text-to-DAUs, and train a highly capable speechLM using these insights.
\end{abstract}

\section{Introduction}
\label{sec:introduction}
% \raul{TODO: ensure all acronyms are defined first time used.}
% \avihu{\begin{enumerate}
%     \item mention that RVQ is infeasible for BPE (2D)
%     % \item ensure it is clear that metrics are calculated on the original space (not BPE space)
%     \item elaborate on the abnormal WER - as in the rebuttal
%     % \item mention Ankit
%     % \item vowel fix
%     % \item explain the addition of 3 special tokens PAD, BOS, EOS
% \end{enumerate}
% }
Representations of language, written and spoken, in the form of discrete units provide the foundation for many language-processing tasks. 
Some historical examples of these inventories have included diphones, phones, and sub-phones (for spoken language) and graphemes and word fragments (for written language). 
For speech tasks, a classical approach has been to use phonetic representations as a link between text and audio since they can encode prior linguistic knowledge and be perceptually distinctive.
More recently, discrete self-supervised representations that we will describe as {\em Discrete Acoustic Units} (DAUs) have provided an alternative intermediate representation that can exploit learning from very large data resources while dispensing with expert knowledge, and are constructed to retain some of the phonetic attributes that can facilitate intelligibility and reproduce prosodic phenomena.
Whether phonetically motivated or self-discovered, these representations play an important role in the pipeline of many
text-to-speech (TTS) systems, and their accurate prediction
from text remains crucial.
% \avihu{better to split paragraph here}

We observe that both phonetic and DAU sequences 
contain redundancy and predictability as a result of a host of constraints (phonotactical, durational, etc.) and are therefore \emph{compressible}. 
The utility of compressing long sequences by grouping their constituents into variable-length tokens has been widely acknowledged in fields like Natural Language Processing (NLP), where the na\"ive approach of operating on raw-character inputs would lead to unnecessarily long sequences, and impose computational constraints in models like Transformers.
% and tokenization algorithms that arrive at vocabularies containing longer textual units are now standard. 
% \avihu{shorten too}
Similar developments, however, have not yet been as widely adopted 
% in the literature
when dealing with acoustic units (with a notable recent exception~\cite{Lajszczak-Cambara:24basetts}) despite the more salient need when working with acoustic signals that operate at a much higher bitrate than text, making the processing of a few minutes of audio difficult due to the quadratic complexity of Transformer models.
% In current setup
% here the need for this is made even more salient by the fact that acoustic signals are very high bit rate: modeling sequences of a few minutes \avihu{maybe we can say that text is a compressed form of audio, so audio is less compressed.
% very high bit rate: modeling sequences of a few minutes \avihu{maybe we can say that text is a compressed form of audio, so audio is less compressed.}
% , or even seconds, \avihu{not true..}
% , which underlie most modern neural architectures and involve operations quadratic in the input length. 

\begin{figure}
    \centering
     \begin{subfigure}[b]{0.3\linewidth}
         \centering
         \includegraphics[width=\textwidth]{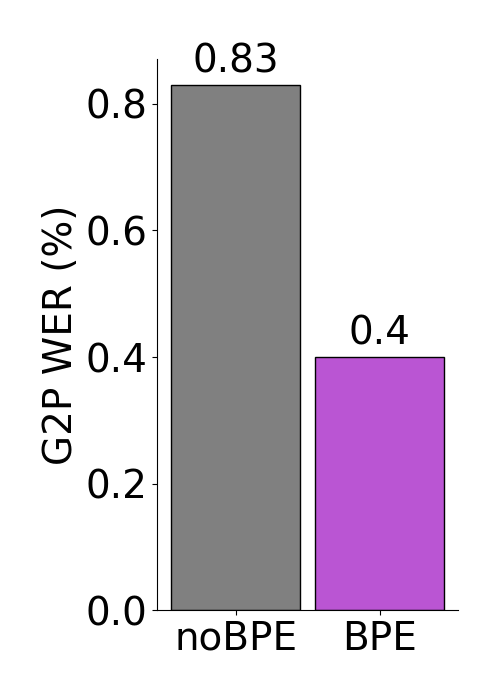}
         \caption{G2P $\shortdownarrow$}
         \label{fig:g2p}
     \end{subfigure}
     \hfill
     \begin{subfigure}[b]{0.3\linewidth}
         \centering
         \includegraphics[width=\textwidth]{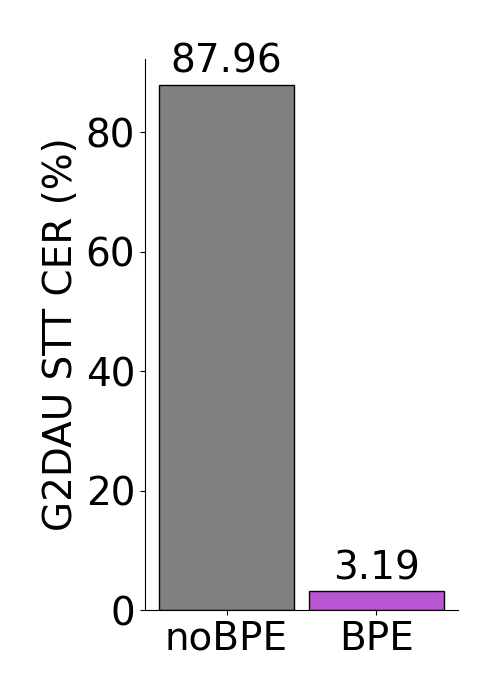}
         \caption{G2DAU $\shortdownarrow$}
         \label{fig:g2dau}
     \end{subfigure}
     \hfill
     \begin{subfigure}[b]{0.3\linewidth}
         \centering
         \includegraphics[width=\textwidth]{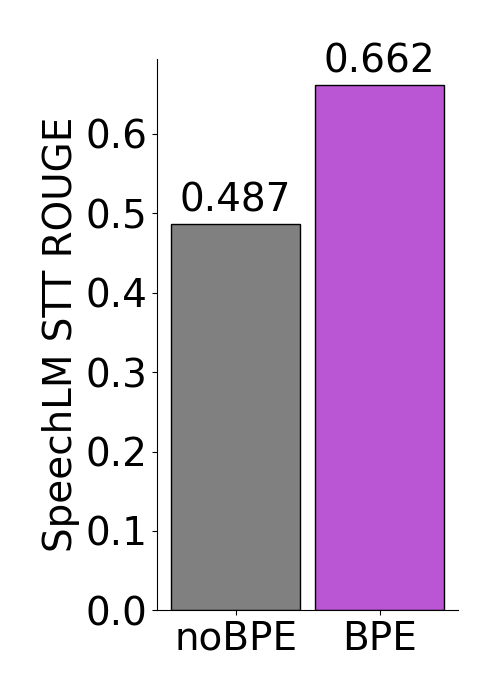}
         \caption{SpeechLM $\shortuparrow$}
         \label{fig:speechlm}
     \end{subfigure}
    \caption{Summarizing the benefits of BPE on DAUs/Phonemes on three tasks. 
    The experimental setup is described in Sec.~\ref{sec:experimental}.}
    \label{fig:summary}
    \vspace{-0.5cm}
\end{figure}

In this paper, we look closely at the tokenization of phonemes and DAUs, and demonstrate that exploiting this during training is beneficial to learning a task both in terms of training and inference speed as well as the final performance.
We document the advantages on three commonly used and important tasks: 
%and which represent novel contributions 
(a) grapheme-to-phoneme (G2P) conversion,
%~\cite{Ploujnikov-Ravanelli:22soundchoice,Dekel-Shechtman:24}, 
(b) prediction of acoustic units from text (G2DAU),
%~\cite{Kharitonov-Vincent:23spear} 
and (c) audio generation using a speech language model (SpeechLM).
%~\cite{Borsos-Marinier:23audiolm,Hassid-Remez:23twist}.  \avihu{do we cite in the intro?}
% as it alleviates imbalance issues present in the distribution of the original data, and thereby frees the model to focus on more challenging aspects of the learning \avihu{too early to discuss imbalance? move later on and add sequence length in AR models}.
To probe this, we adopt the Byte Pair Encoding (BPE) algorithm~\cite{Gage:94,Sennrich-Hadow:16bpe}, a simple yet effective way to derive a new vocabulary by the iterative grouping of frequent pairs of elements, in order to reduce the length of sequences at the expense of creating a larger vocabulary. 
% Despite the simplicity of the approach, 
Fig.~\ref{fig:summary} summarizes the main findings of this BPE
exploration.
To the best of our knowledge, this is the first in-depth 
performance evaluation of BPE, or other tokenization algorithm, that has been carried  
out for acoustic units. 
% In this work, we explore the advantages of tokenization, and demonstrate
% that training with tokenized DAUs yields significant improvements
% in terms of performance, as well as in terms of training and inference speed, 
% We document these advantages within three applications:
% classical grapheme-to-phoneme conversion~\cite{Ploujnikov-Ravanelli:22soundchoice,Dekel-Shechtman:24},  prediction of acoustic units from text~\cite{Kharitonov-Vincent:23spear}, and audio generation via language  models~\cite{Borsos-Marinier:23audiolm,Hassid-Remez:23twist}.
We provide theoretical insights to shed light on the performance benefits of BPE, including the BPE influence on token imbalance and the connection between sequence length and accuracy in autoregressive models.
% issues present in the distribution of the original data, and thereby frees the model to focus on more challenging aspects of the learning \avihu{too early to discuss imbalance? move later on and add sequence length in AR models}.
We hope that these findings will lead to a wider adoption 
% by the community
of tokenization algorithms within models that deal with acoustic or phonetic units.

%upon DAUs and phonemes in the community.
% (parallelling its use in the NLP field) 
%within three tasks that this paper focuses on, %and which represent novel contributions 
%to the field: grapheme-to-phoneme conversion~\cite{Ploujnikov-Ravanelli:22soundchoice,Dekel-Shechtman:24}, prediction of acoustic units from text~\cite{Kharitonov-Vincent:23spear} and audio generation via language models~\cite{Borsos-Marinier:23audiolm,Hassid-Remez:23twist}. 

\subsection{Summary of contributions} 
Our contributions are three fold:
\begin{enumerate}
\item We quantify the compression benefits of applying BPE on discrete audio and phonetic units.
\item We show significant improvements in performance metrics as well as speedups on 
G2P, G2DAU, and SpeechLM tasks.
% \item We demonstrate the applicability of the approach to the end task of spoken language
% generation. 
%\item We connect the BPE impact with mitigating data imbalance and reducing the sequence length in autoregressive models.
\item We show the impact of BPE on mitigating data imbalance and on reducing sequence length in  autoregressive models.
\end{enumerate}

\section{Related Work}
\label{sec:related_work}
{\bf DAUs} are discrete representations of audio signals, usually quantized embeddings from a pre-trained self-supervised speech model (such as HuBERT~\cite{Hsu-Bolte:21hubert}, Wav2Vec2~\cite{Baevski-Zhou:20wav2vec2}, WavLM~\cite{Chen-Wang:22wavlm}, or Whisper~\cite{Radford-Kim:23whisper}).
Representing continuous-valued high-frequency signals (like speech or audio) with a finite vocabulary of units computed at a much slower rate has led to recent advances in LM techniques when modeling
the resulting signals, and to a fruitful field of 
{\it audio/speech language modeling} (e.g., \cite{Lakhotia-Kharitonov:21genlang}, AudioLM~\cite{Borsos-Marinier:23audiolm},
TWIST~\cite{Hassid-Remez:23twist}, and SpeechLM~\cite{Zhang-Chen:23speechlm}).
As DAUs contain important phonetic and suprasegmental information, they have been used as  
a coarser intermediate representation
when predicting the more fine-grained acoustic tokens of a neural codec in discrete 
TTS systems (like SPEAR~\cite{Kharitonov-Vincent:23spear} and 
% audio-generation models (like AudioLM~\cite{Borsos-Marinier:23audiolm} and 
Soundstorm~\cite{Borsos-Sharifi:23soundstorm}),  thus
assuming the more classical function of conditional phonetic units, albeit in a purely data-driven way\footnote{While the term {\em Semantic Token} has been proposed~\cite{Borsos-Marinier:23audiolm}, we adopt the term DAU to be more neutral about the nature of these representations, and be
consistent with the unit-vs-token nomenclature of Sec.~\ref{ssec:bpe}}.
% be more neutral about the nature of the extracted representation and to 
% be consistent with the unit-vs-token nomenclature
% proposed earlier.
%However, since the time scale and mode of construction of these units suggest that they are closer to an over-complete allophonic-phonetic dictionary augmented with suprasegmental features, and the main findings of this paper could apply to other discrete acoustic inventories, we adopt a more neutral term. 
Such property is also leveraged by other neural (but not discrete) TTS (Tacotron-like) architectures to operate directly on DAU inputs rather than phones~\cite{Garg-Kim:24, Liu-Ling2:23}.
% Examples include~\cite{Garg-Kim:24} and~\cite{Liu-Ling2:23}) operating on HuBERT and wav2vec 2.0 units, respectively \avihu{just cite in previous sentence, no?}.
The phonetic and lower bit-rate properties of DAUs have also made them amenable inputs to a lightweight codec that, once augmented with pitch
and speaker embeddings, is able to resynthesize and manipulate (e.g., voice-convert) 
speech~\cite{Polyak-Adi:21}. 
Finally, DAUs have also been used as a proxy for textual representations in
unsupervised speech-to-speech translation~\cite{Lee-Gong:22,Lee-Chen:22}. 
%DAUs can also stand as a proxy for textual representations,
%as has been done in {\em textless-NLP} applications like unsupervised speech-to-speech translation~\cite{Lee-Gong:22,Lee-Chen:22,Popuri-Chen:22}. \avihu{SpeechLMs are also textless-NLP}

{\bf Tokenization} algorithms have been widely adopted in NLP,
% , as already mentioned, 
with a variety of algorithms proposed,
% examples of these approaches applied to textual signals 
including
{\it WordPiece}~\cite{Schuster-Nakajima:12,Wu-Schuster:16},
sub-word level BPE~\cite{Vaswani-Shazeer:17attention}, 
Unigram
% Language Modeling (ULM) approach for word segmentation
~\cite{Kudo:18}, and Sentence Piece~\cite{Kudo-Richardson:18}, with BPE arguably being the most common tokenization algorithm due to its simplicity.
% these tokenization algorithms \avihu{due to their simplicity}. 
% or variants of it,
% plays an important role within
% these tokenization algorithms \avihu{due to their simplicity}. 
BPE-based tokenization applied to DAUs was only very recently explored for speech synthesis~\cite{Lajszczak-Cambara:24basetts}, though that work does not look to isolate the contribution of tokenization.
Tokenized units have also been exploited within
speech recognition systems, both textually derived (e.g., via BPE as in~\cite{Zeyer-Irie:18}) and acoustically derived (e.g., the ADSM model of~\cite{Zhou-Zeineldeen:21adsm}).
The successive merging of DAUs within BPE leads to a {\em variable-rate} inventory, an idea closely related to work on event-driven audio representations~\cite{Dieleman-Nash:21variablerate, Lisboa-Bellec:24spiking}.
% Slow Autoencoders for speech modeling
% or Event-based Autoencoders for music compression~\cite{}.

%of speech~\cite{Dieleman-Nash:21variablerate}, 
%or by...\avihu{We should ask Slava for more references}
%\avihu{Spiking Music: Audio Compression with Event Based Auto-encoders}
%\avihu{mention Encodec Huffman coding for compression? }

Work in end-to-end TTS has demonstrated the ability of these models
to work directly from textual inputs, bypassing explicit intermediate representations
with character-to-acoustic models~\cite{Ping-Peng:17deepvoice3}.
In practice, however, the superior robustness of phonetic inputs has been documented~\cite{Taylor-Richmond:19}, and many modern state-of-the-art systems continue to rely on phonetic inputs 
and a separate \textbf{G2P} module~\cite{Shen-Pang:18,Wang-Chen:23valle,Shen-Ju:24naturalspeech2}, a fact
that continues to fuel development of modern G2P models like T5-G2P~\cite{Rezackova-Svec:21t5gp2}, 
Soundchoice ~\cite{Ploujnikov-Ravanelli:22soundchoice}, ByT5~\cite{Zhu-Zhang:22byt5}, and LLM2Speech~\cite{Dekel-Shechtman:24}. None of these cited works, however, exploit
the advantages of tokenization for G2P prediction that we propose and demonstrate here.

\section{Methodology}
\label{sec:method}

\subsection{Base Unit Construction}
\label{ssec:inputs}
{\bf DAUs}: We extract DAUs following \cite{Lee-Gong:22} using a pre-trained mHuBERT model, extracting the embeddings from the 11th layer and quantizing them into $K=1000$ clusters using pretrained K-Means centroids\footnote{Pretrained quantizer and vocoder are available at: \\
\url{https://github.com/facebookresearch/fairseq/blob/main/examples/speech\_to\_speech/docs/textless\_s2st\_real\_data.md}}. 
This results in a discrete sequence with a frequency of 50 Hz.
% We add 3 special tokens: for BOS, EOS and PAD.
% \avihu{maybe worth discussing the datasets here too?}
% . \raul{How long are the sequences that get embedded every 20 msecs?}. \avihu{10-20 seconds in ml-librispeech}
% We adopt this frame rate as some works have shows that further reducing the frame rate to 25Hz can lead to information loss\raul{TODO:CITE???}. 
% \avihu{No need for that, right?}
{\bf Phones:} For G2P experiments, we generate phonetic sequences from text using a proprietary rules-based phonetizer from the linguistic analysis front-end of a TTS system. This module uses an inventory of 45 phones (17 vowels and 28 consonants) with 3 levels of lexical stress per syllable. 
We extract unique combinations of vowel phones and lexical stress, and merge these with the set of consonant symbols, to arrive at a final phonetic vocabulary containing 81 units (including pause and word separator). Additionally, we include 3 special tokens (PAD, BOS, EOS) for DAUs and phones, to perform autoregressive modeling.
{\bf Datasets:} 
In G2P experiments, we make use of a random subset of the 
{\em Commmon Crawl}~\cite{CommonCrawl} dataset, consisting of 3M/6K train/validation paragraphs respectively. 
Each paragraph was truncated (on sentence ending) to contain at most 200 words.
In DAU experiments, we use the English subset of the {\em multi-lingual LibriSpeech} corpus~\cite{Pratap-Xu:20libri}, which contains 10M/3.8K train/validation transcribed utterances of length 10--20 seconds.
For the SpeechLM experiments, we included additional training-only data from {\em People's Speech}~\cite{Galvez-Diamos:21peoplespeech}, {\em VoxPopuli}~\cite{Wang-Morgane:21voxpopuli}, and {\em Common Voice}~\cite{Ardila-Branson:20commonvoice} \footnote{We thank Ankit Gupta for preparing these datasets.}.

% \raul{CKLST TODO (from submission portal): 
% \begin{itemize}
% %\item Languages, Audio duration distribution, number of examples, labels distribution (or REF answering these details (?)
% \item Train/dev splits (?)
% % \item REF to datasets from literature (DONE)
% % \item Preprocessing details (and any criteria used to filter examples) (DONE)
% \end{itemize}
% }
\subsection{Byte Pair Encoding}
\label{ssec:bpe}
In terms of nomenclature, henceforth a {\em unit} is an entry in the original inventory whereas a {\em token} is an entry in the inventory derived via a
{\em tokenization} algorithm that produces non-uniform groupings of the original unit set. To derive the token vocabularies, we make use of the standard BPE algorithm.
We start with an original discrete vocabulary $\mathcal{X}$, 
where each sample is a sequence of elements from the vocabulary $x=(x_1 ,..,x_n)  $ s.t. $x_i \in \mathcal{X}$. After applying the algorithm described in Alg.~\ref{alg:bpe}, we obtain a new vocabulary
$\mathcal{Z}$ where $|\mathcal{Z}|>|\mathcal{X}|$ and $x$ can now be represented by $z=(z_1,...,z_k)$ where $k<n$.
% These two equivalent formulations are usually learned by some autoregressive neural model.
Denoting the model by $\Theta$ and the context (e.g. text) by $W$, we get two equivalent formulations of the learning task using a neural autoregressive model, 
modeling the original sequence (Eqn.~\ref{eq:prob-org}) or the BPE-derived sequence (Eqn.~\ref{eq:prob-bpe}):
\begin{equation}
p(x_1,...,x_n | \Theta , W)= \prod_{i=1}^n p(x_i | \Theta, W, x_1,..,x_{i-1}),
\label{eq:prob-org}
\end{equation}
\vspace{-0.1cm}
\begin{equation}
p(z_1,...,z_k | \Theta , W)= \prod_{i=1}^k p(z_i | \Theta, W, z_1,..,z_{i-1}).
\label{eq:prob-bpe}
\end{equation}
\begin{algorithm}
\caption{BPE (Byte Pair Encoding)}
\begin{algorithmic}[1]
\State \textbf{Input:} Sequences, Raw Vocab $\mathcal{X}$, Target Size $k$
\State \textbf{Output:} BPE Vocabulary $\mathcal{Z}$
\State $\mathcal{Z} = \mathcal{X}$
\While{$|\mathcal{Z}|<k$}
    \State Find the most frequent pair of adjacent units $(a, b)$
    \State Merge $(a, b)$ to form a new symbol $ab$
    \State Add $ab$ to $\mathcal{Z}$
    \State Replace all occurrences of $(a, b)$ with $ab$
\EndWhile
% \State Tokenize the text using the final vocabulary
\end{algorithmic}
\label{alg:bpe}
\end{algorithm}

Why might the formulation of Eqn.~\ref{eq:prob-bpe} be better given the equivalency of the learning tasks? As has already been pointed out by other authors~\cite{Gowda-May:20bpe4nlp}, by iteratively merging the most frequent pairs into new tokens, 
%BPE balances the tokens' distribution, and it is known that neural models struggle with skewed data distributions.
BPE balances the tokens' distribution, and it is known that skewed data distributions pose
an obstacle to neural models trained with cross-entropy loss.

To quantify this 
effect, we propose the use of normalized entropy, a balance metric that is invariant to vocabulary size. Given a vocabulary $\mathcal{X}$ with a probability distribution over its elements $D(x) : x\in \mathcal{X}$, and the distribution's entropy
given as $H(D)=-\sum_{x\in\mathcal{X}} D(x)\log_2(D(x))$, we define the {\em normalized entropy} as
\begin{equation}
    0 \le N(D) = \frac{H(D)}{\log_2 (|X|)} \le 1,
    \label{eq:normentropy}
\end{equation}
where a value of 1 corresponds to a perfectly balanced distribution (i.e., a uniform multinomial). Table~\ref{tab:imbalance} illustrates the value of this metric before and after applying BPE to the original phonetic and DAU vocabularies, showing the significant change in balance introduced by BPE.

\begin{table}[th]
  \caption{BPE impact on balancing}
  \label{tab:imbalance}
  \centering
  \begin{tabular}{c c c c}
    \toprule
    Domain & BPE & Vocab Size & Balance metric $N(D)$ \\
    \midrule
    \multirow{2}{*}{Phonetic}    & \xmark     & 84    &   0.797    \\
                & \cmark     & 256   &   \textbf{0.919} \\ \midrule
    \multirow{2}{*}{DAU}         & \xmark     & 1003  &   0.876 \\
                & \cmark     & 2048  &   \textbf{0.944} \\
    \bottomrule
    \end{tabular}
\end{table}
\vspace{-0.3cm}

% then train models to translate text into phonemes / DAUs - making use of the trained tokenizer.

%\subsection{Relations to imbalance}
%
%BPE alleviates imbalance.
%BPE iteratively merges the most frequent pairs into new tokens - which effectively reduces imbalance. 
%Imbalance issues are known to be problemantic to neural models trained with cross-entropy, so it is reasonable that this would improve the final model's performance.
% In G2P - the most frequent token is SEP and it appears 18\% of the time.
% Phoneme/semantic token distribution - before and after.
% In the tokenized phonetic description, with a 2048 vocabulary size, the most frequent token is a variant of the word "the" and its probability is 2\%.
%To verify that BPE reduces imbalance, we use normalized entropy, a balance metric that is invariant to the vocabulary size. 
%Given a vocabulary $\mathcal{X}$ and a probability distribution over its elements $D(x) : x\in \mathcal{X}$. Define entropy as $H(p)=-\sum_{x\in\mathcal{X}} D(x)\log(D(x))$ we define the normalized entropy as
%$$N(p) = \frac{H(p)}{\log_2 (|X|}$$
%value of 1 is a perfectly balanced distribution. higher value means a more balanced distribution, results in table~\ref{tab:imbalance}
% Normalized entropy - phonetics, no BPE: 0.797
% Normalized entropy - phonetics, BPE (256): 0.919
% Normalized entropy - semantic, no BPE: 0.876
% Normalized entropy - semantic, BPE (2048): 0.944
% \avihu{find measurements of imbalance to compare}

\subsection{Sequence length in autoregressive models}
We can also obtain insights into the performance of BPE-tokenized models by noting 
the following. An autoregressive model's error rate accumulates as the sequence gets longer. BPE manages to alleviate this by reducing the sequence length, but it does so while increasing the vocabulary size, effectively making the classification of each individual token harder. The adoption of BPE, therefore, introduces a trade-off between sequence length and token-level accuracy. Consider that in the original vocabulary, a sequence has length $n_1$ with a token error rate of $\epsilon_1$, and, after tokenization, length $n_2$ with a {\em token error rate} of $\epsilon_2$, and let's consider the
``edge case'' of having the model classify every token correctly.  Assuming that the average error rate for every token is the same and independently distributed (which is unrealistic), the probability of such, in the  $i^{th}$ scenario, would be $P(correct)_i= (1-\epsilon_i) ^{n_i}$. We illustrate this difference with some actual values from a G2P experiment where the average original sequence length is 
$n_1=872$ and,  after tokenizing with a vocabulary of 2048, $n_2=300$. The corresponding empirical average errors are found to be $\epsilon_1=0.097\%$ 
and $\epsilon_2=0.14\%$, respectively, which leads to $P(correct)_1=42.94\%$
and $P(correct)_2=65.61\%$. With the tokenized
vocabulary, the probability of this edge case is
substantially higher.

% $$p_1 = (1-\epsilon_1 ) ^ {n_1}$$
% $$p_2 = (1-\epsilon_2 ) ^ {n_2}$$
% to reach $p_1 = p_2$ it follows that
% $$(1-\epsilon_1 ) ^ {n_1} = (1-\epsilon_2 ) ^ {n_2}$$
% assuming an original length of $n_1=1000$ tokens, a reduction of $2$ so $n_2=500$ we get
% $$(1-\epsilon_1 ) ^ {1000} = (1-\epsilon_2 ) ^ {500}$$

%BPE reduces the sequence length while increasing the vocabulary, making each token classification task harder. 
%Therefore, different vocabulary sizes trade off sequence length and token level accuracy.
% , the error tends to accumulate as the generated sequence gets longer. Tokenization algorithms reduce the sequence length while increasing the vocabulary (and hence the probability of an error).
%Both the BPE and non-BPE variants are solving the same problem with two different formulations.
%NON-BPE: we have a token error rate of $\epsilon _ 1$ and sequence length of $n_1$
%BPE case: we have a token error rate of $\epsilon _ 2$ and a sequence length of $n_2$.
%The probability of getting the output exactly right (assuming that the average error rate indeed is identical for every token, which is unrealistic):
%$$P(correct)_i= (1-\epsilon_i) ^{n_i}$$
%Let's plug in some account actual numbers, in the G2P case:
%Non-BPE: 
%$$n_1=872, \epsilon_1=0.000969,P(correct)_1=0.42939$$
%BPE (Vocab 2048): 
%$$n_1=300, \epsilon_1=0.001404, P(correct)_2=0.65606$$

%In Section~\ref{sec:experimental}, we compare these formulations in various tasks.

\section{Experiments}
\label{sec:experimental}
% We now turn to investigate the influence of BPE on compressing phonemes and DAUs.
In the following experiments, we apply BPE  with varying vocabulary sizes
to both DAU and phonemes, and compare the performance with respect to
the original discrete vocabularies by training Transformer models
on the following tasks:\footnote{With phonemes, we apply sub-word level tokenization: we do not merge the phonemes of different words.}
\begin{enumerate}
    \item G2P: Grapheme to Phoneme Prediction   (Sec.~\ref{sec:g2p})
    \item G2DAU: Grapheme to Discrete-Audio-Units (Sec.~\ref{sec:text_to_dau})
    \item SpeechLMs using Discrete-Audio-Units (Sec.~\ref{sec:speechlm})
\end{enumerate}
% \raul{DISCUSS: Speech LM may be an overloaded term. It is a reference to a particular paper/model, and a more generic term for a discrete speech-generation model.}

% \avihu{motivate this model, cause we already have a rule-based system that we try to mimic.}
% \avihu{showcase some real examples in phonetics? run a TTS?}
% \raul{I don't think the use of phonemes needs much motivation, beyond the fact that it is a linguistically motivated description of what is happening in language, and part of the pipeline in many TTS (and other LT) systems, including E2E models. The motivation for using them is *not* we have a rules-based legacy FE that happens to use them.}
% \avihu{let's discuss - replace FE with a calculator. the point is we don't need a neural network to be a calculator, we already have a functioning calculator that works perfectly fine.}

% Possibly - adding additional
% To train the SpeechLMs, we also make use of PeoplesSpeech, VoxPopuli, CommonVoice 11.0

\subsection{Evaluation Metrics}
\subsubsection{BPE Evaluation}
To evaluate BPE compression, we make use of the following metrics. First, the {\em reduction in sequence length}, and the relative increase in the number of bits needed to represent the vocabulary, are given by: 
\begin{equation}
    \text{Reduction}=\frac{\hat n}{\hat k}, \quad 
    \text{BitIncrease}=\frac{\log_2(\mathcal{|Z|})}{\log_2(\mathcal{|X|})},
\end{equation}
where $\hat n , \hat k$ denote the average length of the original and BPE sequence, respectively.
% \begin{equation}
% \end{equation}
Using those, we define the {\em compression} achieved by BPE as:
% to represent the reduction in the memory required to store the sequence, as:
\begin{equation}
  \text{Compression}=\frac{\text{Reduction}}{\text{BitIncrease}}=\frac{\hat n\log_2(\mathcal{|X|})}{\hat k\log_2(\mathcal{|Z|})}  
\end{equation}

\subsubsection{Task Evaluation}
\label{ssec:taskeval}

% \raul{I think we should avoid using hyperlinks with href, as per the template guidelines.}
%\subsubsection{Hyperlinks}
%For technical reasons, the proceedings editor will strip all active links from the papers during processing. URLs can be included in your paper, if written in full, e.g., \url{https://www.interspeech2023.org/call-for-papers}. The text must be all black. Please make sure that they are legible  when printed on paper.

For {\bf G2P} we follow standard practice and report Word Error rate (WER) 
%measuring the fraction of words that the model got wrong
\footnote{All metrics are computed in the original vocabulary.}. 
For the {\bf G2DAU} task, as there are various possible options for a correct DAU translation, we follow the work in SPEAR TTS~\cite{Kharitonov-Vincent:23spear} and calculate the Character Error Rate (CER) obtained with an external Speech-to-Text (STT) system. Specifically, we synthesize the DAU tokens using the pre-trained vocoder described in~\cite{Lee-Gong:22}
%\footnote{\href{https://github.com/facebookresearch/fairseq/blob/main/\newline examples/speech\_to\_speech/docs/textless\_s2st\_real\_data.md}{Pretrained quantizer and vocoder are available here}}, 
, and apply STT using Whisper-large-v3 to translate the audio back to text, and
calculate and report CER between the original text and the STT output. 

Finally, for the {\bf SpeechLM} task, we evaluate the generated audio
against a selected reference (details in Sec.~\ref{sec:speechlm}) by transcribing it with
STT (as above) and computing CER plus
two other established metrics from NLP for text comparison: BLEU (for machine translation) and ROUGE (for
summarization). 
We additionally evaluate the quality of the generation by scoring the STT transcripts with the Mixtral8x7B LM~\cite{jiang2024mixtral}.
Given a prompt and two continuations, the LM is asked to select which continuation is better, given the following evaluation criteria: The continuation should (a) be not too short (at least a sentence long), (b) not contain repetitions, (c) be a sensible continuation, and (d) be creative. Each comparison was done twice, replacing the transcripts' ordering.
% We additionally evaluate the quality of the generation by scoring the STT transcripts with an independent LM, in our case 
% the Mistral7B model~\cite{Mistral7B}, and reporting average negative 
% log-likelihood (NLL). 
To allow for a qualitative impression of quality and prosody, we provide a page with samples synthesized using the pretrained vocoder\footnote{Sample page is available here: 
%\url{https://s3.us-south.objectstorage.softlayer.net/zk-wav-data/Webpages/BPETokenization_IS2024/index.html}}.
\url{https://ibm.biz/BdmLCb}}.

Finally, we measure speedup gains by reporting the relative %speedup comparing to the non-BPE model factor of BPE-tokenized models with respect to the base model, based on the
increase in the number of batches per second, compared to the baseline model that does not apply BPE. We ensure all computing resources are identical within a set of experiments. Results are reported for the training set, though similar speedups are obtained for inference.

%TODO: Speed=Batches per second. And everything within a single table uses cmparable computing resources, same batch size, etc. 

%Training/Inference speed: measured by the number of batches per second - we can normalize (divide by 2.11) write training speedup. We see similar numbers in inference. \avihu{visualize convergence speed?}

% \raul{I think a plot of loss vs training time (however expressed) is a good way to present the convergence differences.}

%In SpeechLMs, we train models on unconditional DAU autoregressive prediction.
%We follow \cite{Hassid-Remez:23twist} and report sWUGGY and sBLIMP \cite{Nguyen-Seyssel:20}
%\avihu{TODO}
%sWUGGY - two options, should have a higher probability for the right option.
%sBLIMP - gramatical vs non grammatical sentences.

\subsection{Model Training}
\label{ssec:training}

We use the {\em T5-small} Encoder-Decoder architecture (75M parameters) for the G2P experiments, and 
{\em T5-base} (280M parameters) for the G2DAU experiments~\cite{Raffel-Shazeer:20t5}.
All models are optimized using AdamW~\cite{Loshchilov-Hutter:19adamw} with
a batch size of 32, and weight decay of $0.1$. The learning rate
is linearly increased to \num{1e-4} over 10k warm-up steps, and annealed using a
cosine schedule over 400k iterations. 
For G2P we train with two V100 GPUs, and for G2DAU  with two A100 GPUs.
All weights are initialized using Xavier initialization, and we use greedy autoregressive decoding for inference.
% https://proceedings.mlr.press/v9/glorot10a/glorot10a.pdf?hc_location=ufi
For SpeechLMs, we train a decoder-only model based on the LLaMA~\cite{Touvron-Lavril:23llama} architecture, with 24 layers of dimensionality 1024, 16 heads per layer, and feed-forward network of size 4096 (400M parameters). 
Each model is trained with four A100 GPUs with a batch of 64 samples, for 1M iterations. 
During inference, we sample the next token with a temperature of 1, sampling from the top 20 tokens, using beam search with 4 beams, and a repetition penalty of 1.2.

%\raul{CKLST TODO: 
%\begin{itemize}
    %\item Details on model architecture (DONE?) 
    %\item Parameter initialization (DONE)
    %\item Model size in num of params. (Done)
    %\item Computing infrastructure (DONE)
    % \item Avg running time for training/inference (DONE)
%\end{itemize}
%}

\begin{table}[th]
  \caption{Results of applying BPE to G2P. (First row indicates the original vocabulary.)}
  \label{tab:bpe_g2p}
  \centering
  \begin{tabular}{c c c c c c}
    \toprule
    Vocab & $\shortuparrow$Reduction  & $\shortuparrow$Compression & $\shortdownarrow$WER \% & $\shortuparrow$Speed \\
    \midrule
%   84   & 1                & 1             & 0.83     & 2.35 \\
    84   & 1                & 1             & 0.83     & 1x \\
%    256  & 1.69             & 1.35          & 0.69     & 2.98 \\
    256  & 1.69             & 1.35          & 0.69     & 1.27x \\
%    512  & 2.03             & 1.44          & 0.58     & 3.48 \\
    512  & 2.03             & 1.44          & 0.58     & 1.48x \\
%    1024 & 2.43             & 1.55          & \textbf{0.40}     & 3.97 \\
    1024 & 2.43             & 1.55          & \textbf{0.40}     & 1.69x  \\
%    2048 & 2.90             & 1.69          & 0.46     & \textbf{4.12} \\
    2048 & \textbf{2.90}             & \textbf{1.69}          & 0.46     & \textbf{1.75x} \\
    % 4096 & 3.43             & 1.82          & -        & 3.82 \\
    % 8192 & \textbf{3.97}    & \textbf{1.95} & -        & 3.48 \\
    \bottomrule
  \end{tabular}
\end{table}
\vspace{-0.3cm}

\subsection{Task 1: Grapheme to Phonemes}
\label{sec:g2p}
Results in Table~\ref{tab:bpe_g2p} show that BPE exploits redundancy in the 
raw phonetic representation (row 1) in order to compress (rows 2-5).
Training and inference time are shorter with BPE tokenization, and accuracy is superior.

% This argument is already made in the writing for Task2, so removing from here.
%The sequence reduction rate is not "worth" the larger matrix multiplication in prediction. 
%Increasing vocab leads to increased memory too, cause of the logits matrix of:
%batch size x seq-len x vocab size. 

\begin{table}[th]
  \caption{Results of applying BPE to G2DAU. (First row indicates the original vocabulary).
  STT CER is also reported for the raw audio (Orig)  and the vocoder reconstruction of the ground truth DAUs (Recon).}
  \label{tab:bpe_dau}
  \centering
  \begin{tabular}{c c c c c}
    \toprule
    Vocab & $\shortuparrow$Reduction & $\shortuparrow$Compression & $\shortdownarrow$CER \% & $\shortuparrow$Speed \\
    \midrule
%    1003   & 1    & 1    & 87.96 & 1.97 \\
    1003   & 1    & 1    & 87.96 & 1x \\
%    2048  & 1.89 & 1.71 & 9.94 & 3.84 \\
    2048  & 1.89 & 1.71 & 9.94 & 1.95x \\
%   4096  & 2.39 & 1.98 & 5.12 & 4.49 \\
    4096  & 2.39 & 1.98 & 5.12 & 2.28x \\
%    8192  & 2.81 & 2.15 & 3.32 & \textbf{5.02} \\
    8192  & 2.81 & 2.15 & 3.32 & \textbf{2.55x} \\
%    16384 & 3.20 & 2.27 & \textbf{3.19} & 4.62 \\
    16384 & \textbf{3.20} & \textbf{2.27} & \textbf{3.19} & 2.35x \\
    \midrule
    Orig & - & -  & 2.04 & - \\ 
    Recon & - & - & 7.41 & -  \\ 
    % 32768 & \textbf{3.57} & \textbf{2.37} & - & - \\
    \bottomrule
  \end{tabular}
\end{table}
\vspace{-0.3cm}

\subsection{Task 2 : Graphemes to Discrete Acoustic Units}
\label{sec:text_to_dau}

Results in Table~\ref{tab:bpe_dau} show significant benefits on the CER 
when applying BPE, with the reduction in sequence length greatly influencing the training and inference speed. One possible explanation for why the advantages of tokenization are more apparent here than for the G2P task
lies in the structure of the DAUs: The scale at which they are extracted leads to sequences with frequent repetitions (e.g., {\em aaabbbbcccccdd}). 
This fact can be exploited during teacher-forcing training by a simple heuristic that copies the prediction of the previous tokens, reducing the next-token-prediction loss. 
This, however, creates a mismatch with respect to 
the true-inference condition. BPE reduces repetition, thereby mitigating 
this exposure bias, and bringing training and inference closer to each other.
Note that the model using the largest vocabulary size (16384) is {\em not} the fastest. At some point, the sequence-reduction rate no longer compensates for the larger matrix multiplication in prediction. 
Very large vocabularies also increase memory consumption, due to the logits matrix of size $ batch\_size\times seq\_len \times vocab\_size$. 
Note also the high CER in the non-BPE case. To allow for fair comparisons, we ensured all models shared the same training conditions, and, after 400k iterations, this model had not yet reached the high-enough accuracy that AR models require for stable next-token prediction, leading to nonsense sequences and the CER rate reported. This was a stable finding verified across various experiments.

%One possible explanation for that is the structure of the semantic tokens:
%"aaabbbbcccccddabc"
%when it is frequent that the last token is going to repeat, the model might reduce the training loss by predicting the previous token, but this is ineffective when doing generation.

%Why vocab=16384 last isn't the fastest? At some point, the sequence reduction rate is not "worth" the larger matrix multiplication in prediction. 
%Very large vocabs are also increasing memory consumption, due to the logits matrix of:
%batch size x seq-len x vocab size. 

\begin{table}[th]
  \caption{Speech-generation metrics without (row 1) and with (row 2) BPE tokenization.}
  \label{tab:bpe_speechlm}
  \centering
  \begin{tabular}{c c c c c}
    \toprule
    %Vocab & STT CER \% & STT BLEU \% & STT ROUGE\% & Speed\\
    Vocab & $\shortdownarrow$CER \% & $\shortuparrow$BLEU & $\shortuparrow$ROUGE & $\shortuparrow$Speed\\
    \midrule
    \bottomrule
%    1003 & 67.04 & 13.40 & 48.69 & 1.21\\
%    16384 & \textbf{63.02} & 17.25 & 66.17 & \textbf{1.97} \\
    1003 & 67.04 & 0.134 & 0.487 & 1x\\
    16384 & \textbf{63.02} & \textbf{0.172} & \textbf{0.662} & \textbf{1.63x} \\
    \bottomrule
  \end{tabular}
\end{table}
\vspace{-0.3cm}

\subsection{Task 3: SpeechLMs}
\label{sec:speechlm}

In this task we explore the generation of audio without any textual supervision by a speech LM that takes in an audio prompt and {\em continues} the audio. To do this, we segment the first 4 seconds of an audio file to act as the prompt, and withhold the rest of it as the {\em Ground-Truth Continuation} (GTC).
We then evaluate the hypothesized completions of the input prompt by applying the same STT protocol described in section~\ref{ssec:taskeval} and computing CER, BLEU and ROUGE against the GTC reference (Table~\ref{tab:bpe_speechlm}).
It is important to note that there is not a unique way to complete an initial prompt, and that the GTC is but one valid outcome. 
However, given the difficulty of evaluating reference-free speech generation, we can take 
closeness to the GTC to be, on average, one measure of the fitness of
the hypotheses. 
The results in Table~\ref{tab:bpe_speechlm} suggest that the BPE SpeechLM results in better performance than the non-BPE SpeechLM. 
% \raul{Moving this to the task evaluation discussion ssec where all the other
% metrics are introduced and discussed.}
%In addition to CER, we also deploy several metrics established
%in NLP for text comparison: BLEU (for machine translation) and ROUGE (for
%summarization). We also evaluate the quality of the audio completions by scoring %their STT transcriptions with an independent LM, in our case 
%the Mistral7B model \avihu{CITE}, and reporting average negative 
%log-likelihood (NLL) in Table~\ref{tab:nll_speechlm}. 

% Finally, Table~\ref{tab:nll_speechlm} compares the NLL (scored with the
% Mistral7B model) for both completions against GTC and prompt.
% Note that this metric favors longer sequences, since more context reduces token-level uncertainty. Indeed, the NLL for
% the (short) prompt is the worst.
Additionally, we evaluate the quality of the continuations using an LLM as a judge~\cite{zheng2024judging}. 
Given a textual prompt and two continuations, the LM judges which continuation is better (see Sec.\ref{ssec:taskeval}). 
Table~\ref{tab:nll_speechlm} shows the BPE variant is preferable over the non-BPE, and that both still lag behind the GTC.
% We also plot the average number of characters in each model, and indeed the non-BPE model tends to product silence, as it is more common in
% In this specific evaluation, this 
% It is worth mentioning 
% However, that NLL still captures the
% quality of the generation is corroborated by the fact that the GTC
% still obtains the best performance. \avihu{well, it's length is the longest on average}
% \raul{TODO: What to do about this? What can we say here given the length caveats? Are we choosing to leave NLL in?}
% \begin{table}[th]
%   \caption{NLL vs Sequence Length for Various Audio Waveforms: Prompt, Continuations without and with BPE, and GTC.}
%   \label{tab:nll_speechlm}
%   \centering
%   \begin{tabular}{c c | c c | c }
%     \toprule
%      - & Prompt & No-BPE &  With-BPE & GTC   \\
%     \midrule
%     \bottomrule
%     NLL & 5.89 & 5.11  &  4.32 &  4.03 \\
%     Seq L & 50.18 &  96.7 &  174.59 & 207.18 \\
%     \bottomrule
%   \end{tabular}
% \end{table}
\begin{table}[th]
  \caption{LLM-as-a-Judge evaluation results. }
  \label{tab:nll_speechlm}
  \centering
  \begin{tabular}{c c c c}
    \toprule
    A & B & Prefer A \% & Prefer B \%  \\
    \midrule
    BPE & non-BPE & \textbf{69.52} &  30.48 \\
    GTC & BPE & \textbf{77.34} &  22.66 \\
    GTC & non-BPE & \textbf{93.89} &  6.11 \\
    \bottomrule
  \end{tabular}
\end{table}
\vspace{-0.3cm}

%Speech LMs from raw audio, without any textual supervision. 
%We create samples by taking a prompt of X seconds out of the audio, the rest is predicted by the model.
%To evaluate those, we apply the same STT protocol in the section above, and compare the generated text continuations with the GT continuation. 
%While there are many ways to continue the audio prompt, on average, the continuation should be close to the GT.
%We report several metrics commonly used in NLP for text comparisons.
%BLEU \avihu{cite} was suggested in machine translation.
%ROUGE \avihu{cite} is common for summarization. 
%We also measure the average negative log-likelihood of the texts using the Mistral7B model \avihu{cite} 
%Note that this metric favors longer sequences and indeed the prompt NLL is low
%Orig NLL: 4.030222
%BPE NLL: 4.318580
%NON-BPE NLL: 5.109764
%Prompt NLL: 5.893131
%A sanity check is the original avg NLL is the lowest.
%Lengths:
%orig average length is 207.176
%prompt average length is 50.185
%no-bpe-pred average length is 96.700
%bpe-pred average length is 174.593

% {'bpe_bleu': tensor(0.1751),
%  'bpe_cer': tensor(0.6304),
%  'bpe_rouge': tensor(0.6385),
%  'no_bpe_bleu': tensor(0.1319),
%  'no_bpe_cer': tensor(0.6718),
%  'no_bpe_rouge': tensor(0.4688)}

\section{Conclusions}
\label{sec:conclusions}
%\raul{Any limitations of the work: (1) We have only evaluated the "linguistic" component of the generated audio (via STT proxy) and ignored acoustic quality (but this was never the point of this paper, so maybe not bring this up? DISCUSS)}
In this paper we have investigated the effect of tokenizing inventories of
phonetic and discrete audio representations via the BPE algorithm in various tasks that deploy them: their
respective prediction from graphemes, and their use within speech-generation models.
By quantifying the trade-off between reducing sequence length and increasing
vocabulary size for these inventories, and by demonstrating that exploiting BPE consistently outweighs the choice of not including it, both in terms of efficiency and performance, it is our recommendation and hope that practitioners in the field will adopt this or similar tokenization approaches going forward.
Future extensions of this work include investigating the effect of using predicted BPE tokens directly into a full TTS system (without unpacking them into their constituent units), 
and reconciling the variable-rate nature of these tokens with
the constant-frame outputs typically generated by TTS architectures. 
%We have demonstrated that exploiting BPE within algorithms that model discrete phonetic and speech tokens consistently outweighs the choice of not including it, both in terms of efficiency and performance, and it is our recommendation and hope that practitioners in the field will adopt this approach going forward.
%\raul{Isn't there a lot of work already on BPE in STT? We cite some already.}
%\avihu{STT on BPE tokens.}
\bibliographystyle{IEEEtran}
\bibliography{discreteaudio,g2p,nlp,stt,ttse2e}

\end{document}